\documentclass[aps,pra,preprint,superscriptaddress,longbibliography]{revtex4-2}

\usepackage{amsmath}
\usepackage{amsthm}
\usepackage{amsfonts}
\usepackage{amssymb}
\usepackage{xcolor,graphicx}
\usepackage{tikz}
\usepackage{color}
\usepackage[pdftex]{hyperref} 
\usepackage{longtable}
\usepackage{array}
\usepackage{dsfont}

\begin{document}

	\title{Total Angular Momentum Coherent State Fields}

    \author{D. Aguirre-Olivas}
    \email[e-mail: ]{dilia.aguirre@inaoe.mx}
    \affiliation{Instituto Nacional de Astrof\'isica, \'Optica y Electr\'onica, Tonantzintla, Puebla 72840, Mexico.}

    \author{G. Mellado-Villase\~nor}
    \email[e-mail: ]{gmelladov@inaoe.mx}
    \affiliation{Instituto Nacional de Astrof\'isica, \'Optica y Electr\'onica, Tonantzintla, Puebla 72840, Mexico.}

    \author{Benjamin Perez-Garcia}
	\email[e-mail: ]{b.pegar@tec.mx}
	\affiliation{Photonics and Mathematical Optics Group, Tecnologico de Monterrey, Monterrey 64849, Mexico.}

    \author{B.~M. Rodr\'iguez-Lara}
	\email[e-mail: ]{blas.rodriguez@gmail.com}
	\affiliation{Universidad Polit\'ecnica Metropolitana de Hidalgo, Tolcayuca, Hidalgo 43860, Mexico.}

	\date{\today}
	
	\begin{abstract}
    Structured light fields exploit spin and orbital angular momentum for precision manipulation, advanced imaging, and high-capacity communication. 
    Orbital angular momentum coherent state beams interpolate between Hermite– and Laguerre–Gaussian beams, enabling continuous spatial control. 
    We introduce a symmetry-based framework for joint control of polarization and spatial structure under the shared $su(2)$ Lie algebra of spin and orbital angular momentum.
    Within this structure, we construct total angular momentum fields as superpositions of circular polarization and Laguerre–Gaussian beams, and define their $su(2)$ coherent states within fixed-angular-momentum subspaces. 
    A single complex parameter controls both polarization and spatial degrees of freedom, enabling continuous, symmetry-preserving tuning.
	\end{abstract}
	
	\maketitle
	\newpage


\section{Introduction} 
\label{sec:Sec1}

Optical fields carry two forms of angular momentum: spin angular momentum (SAM), associated with polarization~\cite{Beth1936}, and orbital angular momentum (OAM), associated with phase structure~\cite{Allen1992, Barnett1994,FrankeArnold2008}. 
SAM is intrinsic and independent of the reference axis~\cite{Andrews2013}, with left- and right-circular polarizations carrying $\pm \hbar$ per photon~\cite{Das2024}, while linear polarization carries none. 
OAM may be intrinsic or extrinsic~\cite{ONeil2002}. 
Laguerre–Gaussian beams (LGBs) carry intrinsic OAM $\hbar \ell$ per photon with integer topological charge $\ell$~\cite{Allen1999}. 
Symmetric apodization preserves intrinsic OAM~\cite{Berry1998}, while asymmetric induces extrinsic contributions~\cite{ONeil2002}. 
Total angular momentum (TAM) combines both SAM and OAM.

Structured light fields exhibit spatially varying polarization, amplitude, and phase, carrying both SAM and OAM~\cite{Yao2011,Zhan2009,Padgett2017,RosalesGuzman2018}. 
Polarization–spatial coupling arises naturally in them~\cite{VolkeSepulveda2006}, enabling spin-dependent beam shifts~\cite{Bliokh2015a}, spin-to-orbit conversion~\cite{Bliokh2015b}, classical entanglement~\cite{Karimi2015}, and coupled intensity–angular momentum moments~\cite{Volyar2024}. 
Geometric visualizations such as the Poincar\'e sphere~\cite{Naidoo2016,Fu2021,Yu2023} and Majorana constellation~\cite{TorresLeal2024} capture these couplings. 
Structured fields generated using spiral phase plates~\cite{Beijersbergen1994}, $q$- and $j$-plates~\cite{Marrucci2006,Devlin2017}, metasurfaces~\cite{Dorrah2021}, optically active materials~\cite{Karmakar2024}, and phase-only modulators~\cite{Gabriel2018,Gabriel2021,Moreno2025} enable independent or joint control over polarization, amplitude, and phase. 
Recent advances demonstrated cylindrical~\cite{Zhan2009,Chen2014,Khonina2018}, helico-conical~\cite{MedinaSegura2023}, and hybrid Hermite–Laguerre–Gaussian beams~\cite{MedinaSegura2025} with tailored angular momentum structure, as well as the generation, filtering, and processing of TAM eigenstates~\cite{Shen2019,Li2023}, driving applications in optical trapping~\cite{Padgett2011,Moradi2019,Guzman2022}, high-resolution imaging~\cite{Abouraddy2006,Bautista2017}, and classical and quantum communication~\cite{Nagali2010,Bozinovic2013,Willner2015,Milione2015,Ndagano2018,Ruffato2019,Zhu2021}.

Here, we construct total angular momentum coherent state (TAMCS) fields, extending our recent orbital angular momentum coherent state (OAMCS) beams that interpolate between HGBs and LGBs~\cite{Aguirre2025}. 
Circular polarization and LGBs form orthonormal bases for SAM and OAM. 
From them, we build TAM fields as Clebsch–Gordan superpositions of spin and orbital components, identifying radial and azimuthal polarization modes within fixed-TAM subspaces. 
We then define TAMCS fields as $su(2)$ coherent superpositions within each TAM subspace, where a single complex parameter jointly controls polarization and spatial structure.

\section{Orthonormal LGBs}
\label{sec:Sec2}

We construct our orthonormal basis for structured light fields as the product of polarization and spatial degrees of freedom (DoFs). 
For the polarization DoF, we use left- and right-circular states.
For the spatial DoF, we use LGBs,
\begin{align}
    \Psi_{p,\ell}(r, \theta, z) = \frac{\sqrt{2}}{w(z)} e^{- \frac{i k r^2}{2 R(z)}} e^{i (2 p + \vert \ell \vert + 1) \zeta(z)} \, \psi_{p,\ell}(\rho, \theta),
\end{align}
with beam width $w(z) = w_0 \sqrt{1 + (z/z_R)^2}$, curvature radius $R(z) = z [1 + (z_R/z)^2]$, and Gouy phase $\zeta(z) = \tan^{-1}(z/z_R)$.
We use the waist radius $w_0 = \sqrt{\lambda z_R / \pi}$, Rayleigh range $z_R = \pi w_0^2 / \lambda$, wavenumber $k = 2\pi / \lambda$, and wavelength $\lambda$.
Each beam includes a transverse profile defined by the Laguerre–Gauss modes,
\begin{align}
    \psi_{p, \ell}(\rho, \phi) 
    = (-1)^{p} \sqrt{\frac{p!}{\pi (p+\vert \ell \vert)!}} \, \rho^{\vert \ell \vert} e^{- \frac{1}{2} \rho^{2}} \mathrm{L}_{p}^{\vert \ell \vert} (\rho^{2}) e^{i \ell \phi},
\end{align}
with scaled radial coordinate $\rho = \sqrt{2}r/w(z)$ and azimuthal angle $\phi$. 
The indices $p$ and $\ell$ define the number of radial and angular nodes, determining the amplitude and phase structure of each beam.

\section{Total Angular Momentum Light Fields}
\label{sec:Sec3}

We construct light fields with well-defined TAM by coupling the polarization and spatial DoFs that carry SAM and OAM, respectively. 
We map circular polarization to the spin-$1/2$ basis $\vert s, m_{s} \rangle$ with $s = 1/2$ and $m_{s} = \pm 1/2$, identifying left-circular polarization with $\vert 1/2, 1/2 \rangle$.
We map LGBs to the Dicke spin-$j$ basis $\vert j, m_{j} \rangle$, with $j = (2p + \vert \ell \vert)/2 = 0, 1/2, 1, \ldots$ and $m_{j} = \ell/2 $ \cite{Morales2024b}. 
This construction exploits the shared $su(2)$ symmetry of SAM and OAM.

Combining $s = 1/2$ and $j = 0, 1/2, 1, \ldots$ using $su(2)$ addition rules produces two TAM subspaces with $J = \vert s - j\vert $ and $J = s + j$, each spanned by $2J + 1$ states labeled by $M = -J, -J + 1, \ldots, J - 1, J$. 
Each TAM state is a SAM-OAM product basis superposition,
\begin{align}
    \vert J, M \rangle = \sum_{m_{s} = - \frac{1}{2}}^{\frac{1}{2}} \sum_{m_{j} = -j}^{j} c_{s, m_{s}; j, m_{j}}^{J, M} \, \vert s, m_{s} \rangle \otimes \vert j, m_{j} \rangle,
\end{align}
with weights given by the Clebsch–Gordan coefficients,
\begin{align}
    c_{s, m_{s}; j, m_{j}}^{J, M} = (-1)^{s - j + M} \sqrt{2J + 1} \left( \begin{array}{ccc} s & j & J \\ m_{s} & m_{j} & -M \end{array} \right),   
\end{align}
expressed via the Wigner $3j$-symbol and computed using the Racah formula \cite{Kobi2008}. 
These coefficients ensure $M = m_{s} + m_{j}$ and enforce the triangle inequality $\vert s - j\vert  \leq J \leq s + j$ \cite{Varshalovich1988}.

We define the optical realization of a TAM state as
\begin{align}
    \bm{E}_{J,M}^{(j)}(\mathbf{r}) =  \sum_{m_{s} = - \frac{1}{2}}^{\frac{1}{2}} \sum_{m_{j} = -j}^{j} c_{s, m_{s}; j, m_{j}}^{J, M} \, \bm{\epsilon}_{m_{s}} \Psi_{j - \vert m_{j} \vert,\, 2 m_{j}}(\mathbf{r}),
\end{align}
where $\bm{\epsilon}_{m_{s}}=\bm{\epsilon}_{\pm 1/2}$ denote left- and right-circular polarization states. 
All LGBs in the superposition share the same total node number $N = 2p + \vert \ell \vert = 2 j$, so they experience identical curvature phase, Gouy phase, and beam scaling.
As a result, our TAM light fields preserve their transverse polarization and spatial structure under paraxial propagation up to a scaling factor.

\begin{figure}[h!]
\centering  \includegraphics[scale=1]{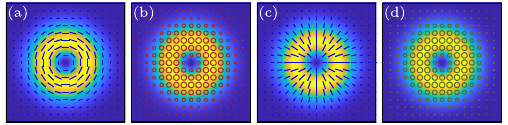}
    \caption{Intensity and polarization distributions for TAM states with $j=1/2$.
    (a) Singlet state with $\left\{ J, M \right\} = \left\{ 0, 0 \right\}$.
    (b)-(d) Triplet states with $J=1$, $M=-1, 0, 1$.
    Vector field colors indicate linear (blue), right- (brown), and left-handed (green) polarizations.}
    \label{fig:Fig1}
\end{figure}

As an example, $j = 1/2$ produces a singlet with $J = 0$ and a triplet with $J = 1$. 
The singlet state with $M = 0$ is
\begin{align}
    \bm{E}_{0,0}^{(\frac{1}{2})}(\mathbf{r}) = \sqrt{\frac{1}{2}} \left[ \bm{\epsilon}_{-\frac{1}{2}} \Psi_{0,1}(\mathbf{r}) - \bm{\epsilon}_{\frac{1}{2}} \Psi_{0,-1}(\mathbf{r}) \right],
\end{align}
with azimuthal polarization, Fig.~\ref{fig:Fig1}(a).  
The triplet states with $M = -1,0,1$ are
\begin{align}
    \begin{aligned}
        \bm{E}_{1,-1}^{(\frac{1}{2})}(\mathbf{r}) &= \bm{\epsilon}_{-\frac{1}{2}} \Psi_{0,-1}(\mathbf{r}), \\
        \bm{E}_{1,0}^{(\frac{1}{2})}(\mathbf{r})  &= \sqrt{\frac{1}{2}} \left[ \bm{\epsilon}_{-\frac{1}{2}} \Psi_{0,1}(\mathbf{r}) + \bm{\epsilon}_{\frac{1}{2}} \Psi_{0,-1}(\mathbf{r}) \right], \\
        \bm{E}_{1,1}^{(\frac{1}{2})}(\mathbf{r})  &= \bm{\epsilon}_{\frac{1}{2}} \Psi_{0,1}(\mathbf{r}),
    \end{aligned}
\end{align}
exhibiting right-circular, radial, and left-circular polarization, respectively, Fig.~\ref{fig:Fig1}(b)–(d).

\section{Azimuthal and Radial Polarization}
\label{sec:Sec4}
\begin{figure}[h!]
\centering  
\includegraphics[scale=1]{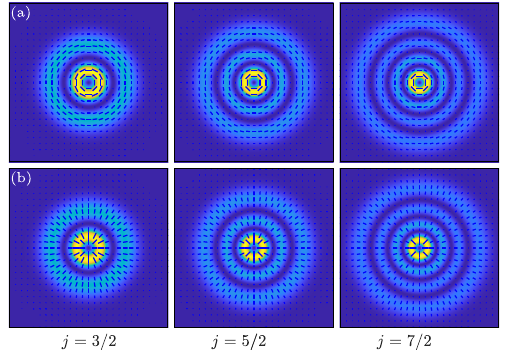}
\caption{Intensity and polarization distributions for (a) azimuthally  and (b) radially polarized TAM fields for $\left\{ J, M \right\} = \left\{ j \pm 1/2, 0 \right\}$ with $j \in \{ 3/2, 5/2, 7/2 \}$.
Vector field colors follow Fig.~\ref{fig:Fig1}.}
\label{fig:Fig2}
\end{figure}

We extend our construction to higher-order OAM components $j = 3/2, 5/2, 7/2, \ldots$, which yield two integer-TAM subspaces with $J = \vert  1/2 \pm j \vert$.
The corresponding $M = 0$ states,
\begin{align}
    \begin{aligned}
        \bm{E}_{j-\frac{1}{2},0}^{(j)}(\mathbf{r}) =&~ \frac{1}{2} \left[ \bm{\epsilon}_{-\frac{1}{2}} \Psi_{j-\frac{1}{2},1}(\mathbf{r}) - \bm{\epsilon}_{\frac{1}{2}} \Psi_{j-\frac{1}{2},-1}(\mathbf{r}) \right],\\
        \bm{E}_{j+\frac{1}{2},0}^{(j)}(\mathbf{r}) =&~ \frac{1}{2} \left[ \bm{\epsilon}_{-\frac{1}{2}} \Psi_{j-\frac{1}{2},1}(\mathbf{r}) + \bm{\epsilon}_{\frac{1}{2}} \Psi_{j-\frac{1}{2},-1}(\mathbf{r}) \right],        
    \end{aligned}
\end{align}
yield azimuthal and radial polarization, respectively, 
\begin{align}
    \begin{aligned}
        \bm{\epsilon}_{-\frac{1}{2}} e^{i \theta} - \bm{\epsilon}_{\frac{1}{2}} e^{-i \theta} &= - 2 i \bm{\epsilon}_{\theta}, \\
        \bm{\epsilon}_{-\frac{1}{2}} e^{i \theta} + \bm{\epsilon}_{\frac{1}{2}} e^{-i \theta} &= 2 \bm{\epsilon}_{r},
    \end{aligned}
\end{align}
where $\bm{\epsilon}_{\theta} = - \bm{\epsilon}_{x} \sin \theta + \bm{\epsilon}_{y} \cos \theta$ and $\bm{\epsilon}_{r} = \bm{\epsilon}_{x} \cos \theta + \bm{\epsilon}_{y} \sin \theta$ are unit vectors for linear polarization in the azimuthal and radial directions, respectively.
Each field pair defines orthonormal polarization states with identical intensity distributions $\vert \Psi_{j - 1, \pm 1}(\mathbf{r}) \vert^{2}$. 
Figure~\ref{fig:Fig2}(a) and Fig.~\ref{fig:Fig2}(b) show fields with azimuthal and radial polarization, respectively, for $\left\{ J, M \right\} = \left\{ j \pm 1/2, 0 \right\}$ with $j \in \{ 3/2, 5/2, 7/2 \}$.

\section{TAMCS}
\label{sec:Sec5}
We previously introduced OAMCS beams that interpolate between HGBs and LGBs, using their $su(2)$ symmetry~\cite{Aguirre2025}. 
Here, we introduce light field analogues to TAMCS,
\begin{align}
    \bm{E}_{J,M,\alpha}^{(j)} (\mathbf{r}) = \sum_{q = -J}^{J} c_{q}(J, M, \alpha) \, \bm{E}_{J,q}^{j}(\mathbf{r}),
\end{align}
where $\alpha = \vert \alpha \vert e^{i \varphi}$ is a complex parameter enabling continuous joint control over polarization and spatial DoFs through the weights,
\begin{align}
    \begin{aligned}
        c_{q}(J, M, \alpha) &= \sqrt{\binom{2J}{J + q} \binom{2J}{J - M}} \, \times \\
        & \times \,_{2}\mathrm{F}_{1}\left(-J - q,\ -J + M;\ -2J;\ \csc^{2} \vert \alpha \vert \right) \, \times \\
        & \times e^{i \varphi (-M - q)} \left(\sec \vert \alpha \vert \right)^{-2J} \left(\tan \vert \alpha \vert \right)^{2J - M + q},
    \end{aligned}
    \label{eq:coeffs}
\end{align}
derived from the optical spatial-mode analogues of quantum coherent states~\cite{Morales2024a, Morales2024b, TorresLeal2024}. 
We use the binomial coefficient $\binom{a}{b} = a!/[b!(a-b)!]$ and the Gauss hypergeometric function ${}_{2}\mathrm{F}_{1}(a,b;c;z)$.
All fields in the superposition belong to the same TAM subspace with modal number $N = j$, so TAMCS fields retain their transverse structure under paraxial propagation up to a scale factor.

\begin{figure}[h!]
\centering  
\includegraphics[scale=1]{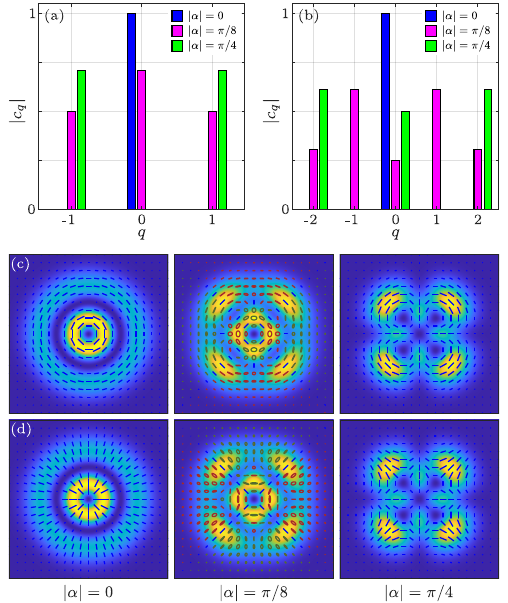}
\caption{(a)-(b) Weight magnitudes $\vert c_q\vert $, (c)-(d) intensity and polarization distribution for TAMCS fields with $\{J, M\} = \{1, 0\}$ with (a),(c) $j=1/2$ and (b),(d) $j=3/2$ for $\varphi = 0$ and $\vert \alpha\vert  \in \{0, \pi/8, \pi/4\}$.
Vector field colors follow Fig.~\ref{fig:Fig1}.}
\label{fig:Fig3}
\end{figure}

We use TAMCS fields with $\{J, M\} = \{1, 0\}$ to illustrate how the amplitude $\vert \alpha\vert $ modulates polarization and spatial structure with $\pi$-periodicity. 
Figures~\ref{fig:Fig3}(a) and \ref{fig:Fig3}(b) show the weight magnitudes $\vert c_q \vert$ for $j = 1/2$ and $j = 3/2$ at fixed phase $\varphi = 0$ and amplitudes $\vert \alpha \vert \in \{0, \pi/8, \pi/4\}$, at $z=0$.
We show the corresponding intensity and polarization distributions in Figs.~\ref{fig:Fig3}(c)–(d).

\begin{figure}[h!]
\centering
\includegraphics[scale=1]{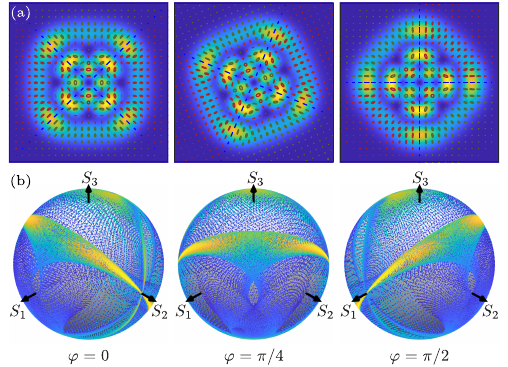}
\caption{(a) Intensity and polarization distribution, and (b) polarization mapped on the Poincar\'e sphere for TAMCS fields with $\{J, M, j\} = \{3, 0, 5/2\}$, $\vert \alpha\vert  = \pi/8$, and $\varphi \in \{0, \pi/4, \pi/2\}$.
Vector field colors follow Fig.~\ref{fig:Fig1}}
\label{fig:Fig4}
\end{figure}

For fixed amplitude $\vert \alpha \vert$, varying the coherent parameter phase $\varphi$ rotates both polarization and spatial structure by $\varphi/2$ around the optical axis with $2\pi$-periodicity.
Figure~\ref{fig:Fig4}(a) shows this effect for $\vert \alpha\vert  = \pi/8$ and phases $\varphi \in \{0, \pi/4, \pi/2\}$, at $z=0$. To characterize polarization, we use the space-resolved Stokes parameters \cite{Goldstein2011},
\begin{align}
    \begin{aligned}
        S_{0}(\mathbf{r}) &= \vert E_{+}(\mathbf{r})\vert ^{2} + \vert E_{-}(\mathbf{r})\vert ^{2},\\
        S_{1}(\mathbf{r}) &= 2\text{Re}\left[ E_{-}(\mathbf{r})^{\ast} E_{+}(\mathbf{r}) \right],\\
        S_{2}(\mathbf{r}) &= -2\text{Im}\left[ E_{-}(\mathbf{r})^{\ast} E_{+}(\mathbf{r}) \right],\\
        S_{3}(\mathbf{r}) &= \vert E_{-}(\mathbf{r})\vert ^{2} - \vert E_{+}(\mathbf{r})\vert ^{2},
    \end{aligned}
\end{align}
in terms of the circular polarization components $E_{\pm}(\mathbf{r})$.
From these, we recover the local polarization ellipse parameters, 
\begin{align}
    A_{\pm}(\mathbf{r}) &= \left[ \frac{S_{0}(\mathbf{r}) \pm \sqrt{S_{1}^{2}(\mathbf{r}) + S_{2}^{2}(\mathbf{r})}}{2}\right]^{1/2},\\
    \chi(\mathbf{r}) &= \frac{1}{2}\arctan \left[ \frac{S_{2}(\mathbf{r})}{S_{1}(\mathbf{r})} \right],
\end{align}
where $A_{\pm}$ are the semi-axes and $\chi$ is the orientation angle.
We map the resulting polarization states for the previous example onto the Poincar\'e sphere in Fig.~\ref{fig:Fig4}(b).
The polarization distribution on the Poincar\'e sphere exhibits rigid rotation by an angle $\varphi/2$ around the $S_{3}$ axis.

\begin{figure}[h!]
\centering
\includegraphics[scale=1]{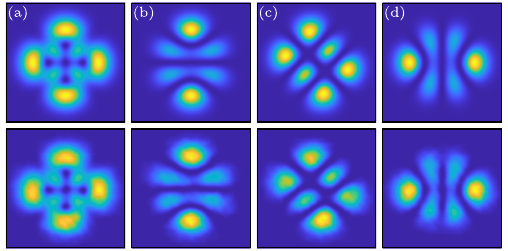}
\caption{Theoretical (upper line) and experimental (lower line) results of a TAMCS field with parameters $\{J, M, j\} = \{1, 0, 3/2\}$, $\vert \alpha\vert  = \pi/4$, and $\varphi = \pi/2$.
Intensity distributions of (a) the TAMCS field and after transmission through a linear polarizer with its transmission axis oriented at (b) $0^\circ$, (c) $45^\circ$, and (d) $90^\circ$, with respect to the horizontal.}
\label{fig:Fig5}
\end{figure}

In Figure~\ref{fig:Fig5}, we illustrate the vectorial nature of the TAMCS field for the parameter set $\{J, M, j\} = \{1, 0, 3/2\}$, $\vert \alpha\vert  = \pi/4$, and $\varphi = \pi/2$. The theoretical and experimental results for this configuration are presented in the upper and lower rows, respectively. The intensity distributions correspond to (a) the TAMCS field and to the field transmitted through a linear polarizer whose transmission axis is set to (b) $0^\circ$, (c) $45^\circ$, and (d) $90^\circ$,  where the angles are referenced to the horizontal axis. 
The experimental results where obtained following the procedure describe in \cite{Gabriel2021}, showing good agreement with the corresponding theoretical simulations.

\section{Conclusion}
\label{sec:Sec6}

In summary, we built an orthonormal basis for structured light fields by pairing circular polarization and LGBs, mapped to spin-$1/2$ and spin-$j$ representations of SAM and OAM, respectively.

We constructed TAM fields as Clebsch–Gordan superpositions of this basis, generating complete subspaces with fixed TAM $J = \vert 1/2 \pm j\vert $ and projection $M = -J, \ldots, J$. 
For $j = 1/2, 3/2, 5/2, \ldots$, the $M = 0$ states yield orthonormal light field pairs with radial and azimuthal polarization and identical intensity profile $\vert \Psi_{j - 1/2, \pm 1}(\mathbf{r})\vert ^2$.

We then introduced TAMCS as $su(2)$ coherent superpositions within each $J$ subspace, governed by a single complex parameter $\alpha = \vert \alpha\vert  e^{i\varphi}$. 
The amplitude $\vert \alpha\vert $ modulates polarization and spatial structure with $\pi$-periodicity, while the phase $\varphi$ induces a rigid rotation around the optical axis with $2\pi$-periodicity. 
These fields preserve both polarization and intensity distributions under paraxial propagation up to a scaling factor.

Our framework enables continuous, symmetry-preserving control of polarization and spatial DoFs through a single complex parameter under $su(2)$ angular momentum symmetry.


\section*{Funding}
Not Applicable. 

\section*{Acknowledgments}
B.~M.~R.~L. acknowledges support and hospitality as an affiliate visiting colleague at the Department of Physics and Astronomy, University of New Mexico.
G.~M.~V. acknowledges SECIHTI Mexico for the postdoctoral grant. 
The authors acknowledge José Miguel Arroyo Hernández from CIT - Optical Workshop (INAOE) for the fabrication of the custom lenses used in this work.

\section*{Disclosures}
The authors declare no conflicts of interest.

\section*{Data Availability Statement}
All the data is available from the corresponding author upon reasonable request.



%

\end{document}